\documentclass[journal,twocolumn]{IEEEtran}
\usepackage{amsfonts}
\usepackage{times}
\usepackage{graphicx}
\usepackage{epstopdf}
\usepackage{latexsym}
\usepackage{dsfont}
\usepackage{amssymb}
\usepackage{amsmath}
\usepackage{cite}
\usepackage{verbatim}

\newtheorem{algorithm}{Algorithm}

\newcommand{\figref}[1]{{Fig.}~\ref{#1}}

\def\bb0{{\mathbb{0}}}

\def\bb{{\mathbf{b}}}

\def\bff{{\mathbf{f}}}

\def\bh{{\mathbf{h}}}

\def\br{{\mathbf{r}}}
\def\bs{{\mathbf{s}}}

\def\bv{{\mathbf{v}}}

\def\bx{{\mathbf{x}}}
\def\by{{\mathbf{y}}}
\def\bz{{\mathbf{z}}}
\def\b0{{\mathbf{0}}}

\def\bA{{\mathbf{A}}}

\def\bF{{\mathbf{F}}}

\def\bI{{\mathbf{I}}}

\def\bP{{\mathbf{P}}}

\def\bR{{\mathbf{R}}}
\def\bS{{\mathbf{S}}}

\def\bW{{\mathbf{W}}}
\def\bX{{\mathbf{X}}}
\def\bY{{\mathbf{Y}}}
\def\bZ{{\mathbf{Z}}}

\def\bbC{{\mathbb{C}}}

\def\bbR{{\mathbb{R}}}

\def\bbZ{{\mathbb{Z}}}

\def\cP{\mathcal{P}}

\def\cS{\mathcal{S}}

\def\sf0{{\mathsf{0}}}

\usepackage{graphicx}
\usepackage{epstopdf}
\usepackage{enumerate}
\usepackage{algorithm}
\usepackage{algorithm,algcompatible,amsmath}
\algnewcommand\INPUT{\item[\textbf{Input:}]}%
\algnewcommand\OUTPUT{\item[\textbf{Output:}]}%
\usepackage{amsmath}
\usepackage{mathrsfs}
\usepackage{float}
\usepackage{color}
\usepackage{makeidx}
\usepackage{bm}
\usepackage{cleveref}
\usepackage{url}
\usepackage{steinmetz}
\usepackage{varwidth}
\usepackage{soul}
\usepackage{bbm}
\usepackage{empheq}
\usepackage{mathtools}
\usepackage{cite}
\usepackage{balance}
\usepackage[scriptsize]{subfigure}

\newcommand{\sref}[1]{{Section}~\ref{#1}}

\def \rm {\mathrm}

\usepackage{amsfonts}
\usepackage{booktabs}
\usepackage{siunitx}

\usepackage{tikz}

\begin{document}
	\title{Sensing Aided OTFS Channel Estimation for \\ Massive MIMO Systems}
	\author{Shuaifeng Jiang and Ahmed Alkhateeb\\
		\textit{School of Electrical, Computer, and Energy Engineering - Arizona State University}
		\\Emails: \textit{\{s.jiang, alkhateeb\}@asu.edu}
	}
	
	\maketitle
	\begin{abstract}
	Orthogonal time frequency space (OTFS) modulation has the potential to enable robust communications in highly-mobile scenarios. Estimating the channels for OTFS systems, however, is associated with high pilot signaling overhead that scales with the maximum delay and Doppler spreads. This becomes particularly challenging for massive MIMO systems where the overhead also scales with the number of antennas. An important observation however is that the delay, Doppler, and angle of departure/arrival information are directly related to the distance, velocity, and direction information of the mobile user and the various scatterers in the environment. With this motivation, we propose to leverage radar sensing to obtain this information about the mobile users and scatterers in the environment and leverage it to aid the OTFS channel estimation in massive MIMO systems. 
	
	As one approach to realize this vision, this paper formulates the OTFS channel estimation problem in massive MIMO systems as a sparse recovery problem and utilizes the radar sensing information to determine the support (locations of the non-zero delay-Doppler taps). The proposed radar sensing aided sparse recovery  algorithm is evaluated based on an accurate 3D ray-tracing  framework with co-existing radar and communication data. The results show that the developed sensing-aided solution consistently outperforms the standard sparse recovery algorithms (that do not leverage radar sensing data) and leads to a significant reduction in the pilot overhead, which highlights a promising direction for OTFS based massive MIMO systems.  
\end{abstract}


\section{Introduction}
Orthogonal time frequency space (OTFS) modulation is a promising approach for achieving robust communication in highly-mobile scenarios.  This is thanks to multiplexing the information bearing data into the nearly-constant channels in the delay-Doppler domain \cite{Hadani2017,Sayeed2021,shen2019channel}. Realizing these gains in massive MIMO systems, however, is challenging. This is mainly due to the high downlink pilot overhead which scales with the maximum delay spread and  the maximum Doppler spread of the channel and with the number of antennas at the transmitter \cite{shen2019channel,guo2020convolutional}. This motivates the  development of novel approaches that enable the OTFS gains in massive MIMO systems, which is the objective of this paper. 

For massive MIMO-OTFS systems, the channels typically experience 3D sparsity in the delay, Doppler, and angle dimensions \cite{shen2019channel,rasheed2020sparse}. The channel sparsity in the delay dimension is due to the limited number of dominant propagation paths compared to the considered delay range, while the sparsity in the Doppler dimension goes back to the small Doppler frequency of the dominant paths compared to the system bandwidth. For the angle dimension, the channel sparsity is a result of the usually small angle of departure (AoD) spread for the propagation paths. Exploiting this channel sparsity in the three dimensions, prior work used different compressive sensing (CS) approaches to reduce the pilot overhead in estimating the OTFS massive MIMO channels \cite{shen2019channel, rasheed2020sparse, zhang20182d, shi2021deterministic}. Despite its reduction, however, this channel acquisition overhead could still be significant for large-scale MIMO systems, especially in scenarios with large delay and Doppler spreads.

\textbf{Contribution:} The delay-Doppler domain channel has a close and direct relation to the position, direction, and velocity of the mobile users and the various scatterers in the surrounding environment. Based on that, in this paper, \textbf{we propose to leverage the radar sensing information about the users and the surrounding environment to aid the OTFS channel estimation in massive MIMO systems.}  This is further motivated by the potential integration  of sensing and communication in future  systems at which the sensing information could potentially be collected with negligible overhead on the wireless communication resources \cite{Demirhan_mgazine_radar,liu2020joint,Demirhan2022,Zhang2021radar,Kumari18}. The contributions of the paper can be summarized as follows. 
\begin{itemize}
	\item Proposing a novel approach that utilizes the radar sensing information at the basestation to facilitate the massive MIMO OTFS channel estimation with a significant reduction in the pilot overhead. 
	\item Developing a sensing framework that infers the delay, Doppler, and the AoD of the communication channel paths using information collected from the radar signals.
	\item Designing an orthogonal matching pursuit (OMP) based algorithm that utilizes the extracted propagation delay, Doppler frequency, and  AoD to improve the  sparse OTFS channel recovery performance.
	\item Developing a new simulation framework with co-existing wireless communication and radar sensing data and adopting it to evaluate the performance of the proposed sensing-aided OTFS channel estimation approach.
\end{itemize}

Simulation results show that the proposed sensing-aided OTFS channel estimation approach consistently outperforms the conventional sparse recovery algorithms. Specifically, the proposed approach can achieve similar channel estimation normalized mean square error (NMSE) performance with $5$dB lower SNR. Further, the proposed approach can lead to more than $50 \%$ reduction in the pilot/channel acquisition overhead without any degradation in the channel estimation NMSE.

	\section{System, Signal,  and Channel Models}\label{System Model and Signal Model}
	In this section, we first present the adopted system model. After that, we introduce the discrete-time signal model and the channel model for the considered MIMO-OTFS systems. Lastly, we introduce the adopted radar signal model.
	\subsection{System Model}
	\begin{table*}[]
		\caption{Notation Adopted in the Paper}
		\label{tbl:variable}
		\centering
		\begin{tabular}{|c|c|}
			\hline
			Variable & Description           \\ \hline
			$M$, $N$        & number of transmit symbols in the delay and Doppler dimensions  \\ \hline
			$m$, $m^\prime$ & indices of delay taps\\ \hline
			$n$, $n^\prime$ & indices of Doppler taps\\ \hline
			$M_p$, $M_g$& Lengths of the pilot and guard symbols in the delay dimension\\ \hline
			$A$ & number of communication transmit antennas at the BS   \\ \hline
			$a$ & index of communication transmit antennas at the BS  \\ \hline
			$B$ & number of radar receive antennas at the BS  \\ \hline
			$b$ & index of radar receive antennas at the BS \\ \hline
			$f_c$, $\Delta_f$, $\lambda$, $d$   & carrier frequency, subcarrier spacing, wavelength, and antenna spacing of the communication system \\ \hline
			$f_0$, $S$, $\lambda_{\mathrm{r}}$, $d_{\mathrm{r}}$  & start frequency, slope, wavelength, and antenna spacing of the FMCW radar system \\ \hline
			$L$    & number of communication propagation paths\\ \hline
			$\alpha_i$, $\tau_i$, $\nu_i$, $\psi_i$ & complex gain, delay, Doppler frequency shift, and AoD of the $i$-th communication path \\ \hline
			$J$    & number of detected radar propagation paths\\ \hline
			$\tau^p_j$, $v_j$, $\theta^p_j$ & delay, Doppler velocity, and AoA of the $j$-th detected radar path \\ \hline
			$q$ & index of discrete time\\ \hline
		\end{tabular}
	\end{table*}
	\begin{figure}[t]
		\centering
		\includegraphics[width=1\linewidth]{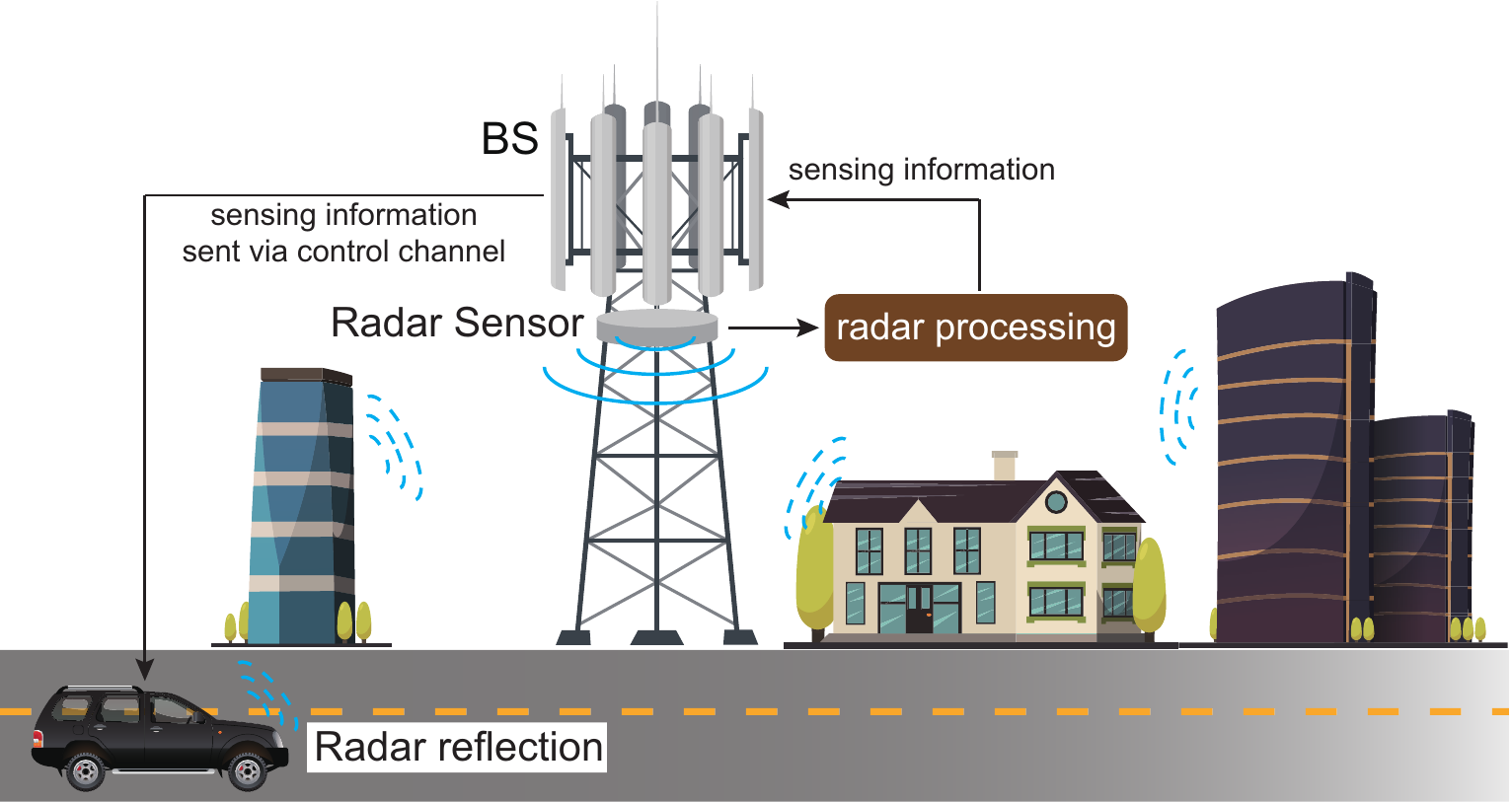}
		\caption{An illustration of the adopted system model. The BS exploits the radar sensing data to aid the MIMO-OTFS channel estimation.}
		\label{fig:sysmodel}
	\end{figure}

	As shown in \figref{fig:sysmodel}, we consider a communication system where a BS serves a single highly mobile UE using a carrier frequency $f_{c}$. The BS employs a uniform linear array (ULA) of $A$ antennas to communicate with the UE. The BS is also equipped with an FMCW radar operating on the start frequency $f_0$. The FMCW radar collects sensing information about the surrounding communication environment, which is then utilized to aid the downlink channel estimation. For simplicity, we assume that the UE is equipped with a single antenna. The proposed sensing-aided OTFS channel estimation in this paper, however, can be extended to systems wtih the multi-antenna UEs. 
	\subsection{OTFS Communication Signal Model}\label{sec:OTFS signal model}
	For the downlink, the BS applies OTFS modulation to prepare the transmitted data. This OTFS modulation is performed as follows. First, $M \times N$ data symbols are arranged into a two-dimensional OTFS frame ${\bX}^{\rm{DD}} \in \bbC^{M \times N}$ in the delay-Doppler domain. After that, the ${\bX}^{\rm{DD}}$ is transformed into the time-frequency domain signal ${\bX}^{\rm{FT}} \in \bbC^{M \times N}$ as
	\begin{align}\label{eq:DD transmit}
	{\bX}^{\rm{FT}} = \bW^{\rm tx} \odot \left({\bF}_{M}{\bX}^{\rm{DD}} {\bF}_{N}^{\rm H}\right),
	\end{align}
	where ${\bf F}_{M}$ and ${\bf F}_N$ denote the $M$-point and $N$-point discrete Fourier transformation (DFT) matrices, and $\bW^{\rm tx}$ is a windowing function\footnote{In the general case where the delay/Doppler values do not belong exactly to the integer delay/Doppler bins, it becomes interesting to optimize the windowing matrix to suppress inter-Doppler interference \cite{wei2021transmitter}.}. The operation $\odot$ denotes the point-wise multiplication. In this paper, we assume that the $\bW^{\rm tx}$ adopts rectangular windowing \cite{raviteja2018interference}. That is, $\bW^{\rm tx}$ is an all-one matrix and therefore can be omitted. The two-dimensional time-frequency domain signal ${\bX}^{\rm{FT}}$ is then converted into a two-dimensional delay-time domain signal ${\bX}^{\rm{DT}}$ as
	\begin{align}
	{\bX}^{\rm{DT}} ={\bF}_{M}^{\rm H}{\bX}^{\rm{FT}}.
	\end{align}
	With ${\bX}^{\rm{DT}} = [\bx_1, \hdots \bx_N]$, each column $\bx_i$ can be regarded as a time-domain OFDM symbol of $M$ subcarriers and ${\bX}^{\rm{DT}}$ consists of $N$ consecutive OFDM symbols. To avoid inter-symbol interference, the cyclic prefix (CP) is added to each OFDM symbol.
	\begin{align}
	\bS = \bA_{\rm{CP}} {\bX}^{\rm{DT}},
	\end{align}
	where $\bA_{\rm{CP}} \in \bbR^{(M+N_{\rm{CP}})\times M}$ is the CP addition matrix. Finally, the discrete-time time-domain baseband transmit signal can be obtained as
	\begin{align} \label{eq:time-domain transmit}
	\bs = \mathrm{vec}(\bS),
	\end{align}
	where $\bs \in \bbR^{(M+N_{\rm{CP}})N \times 1}$ is the concatenation of all the columns in $\bS$ with $\mathrm{vec}(\cdot)$ representing the vectorization operation.
		\begin{figure}[t]
		\centering
		\includegraphics[width=1\linewidth]{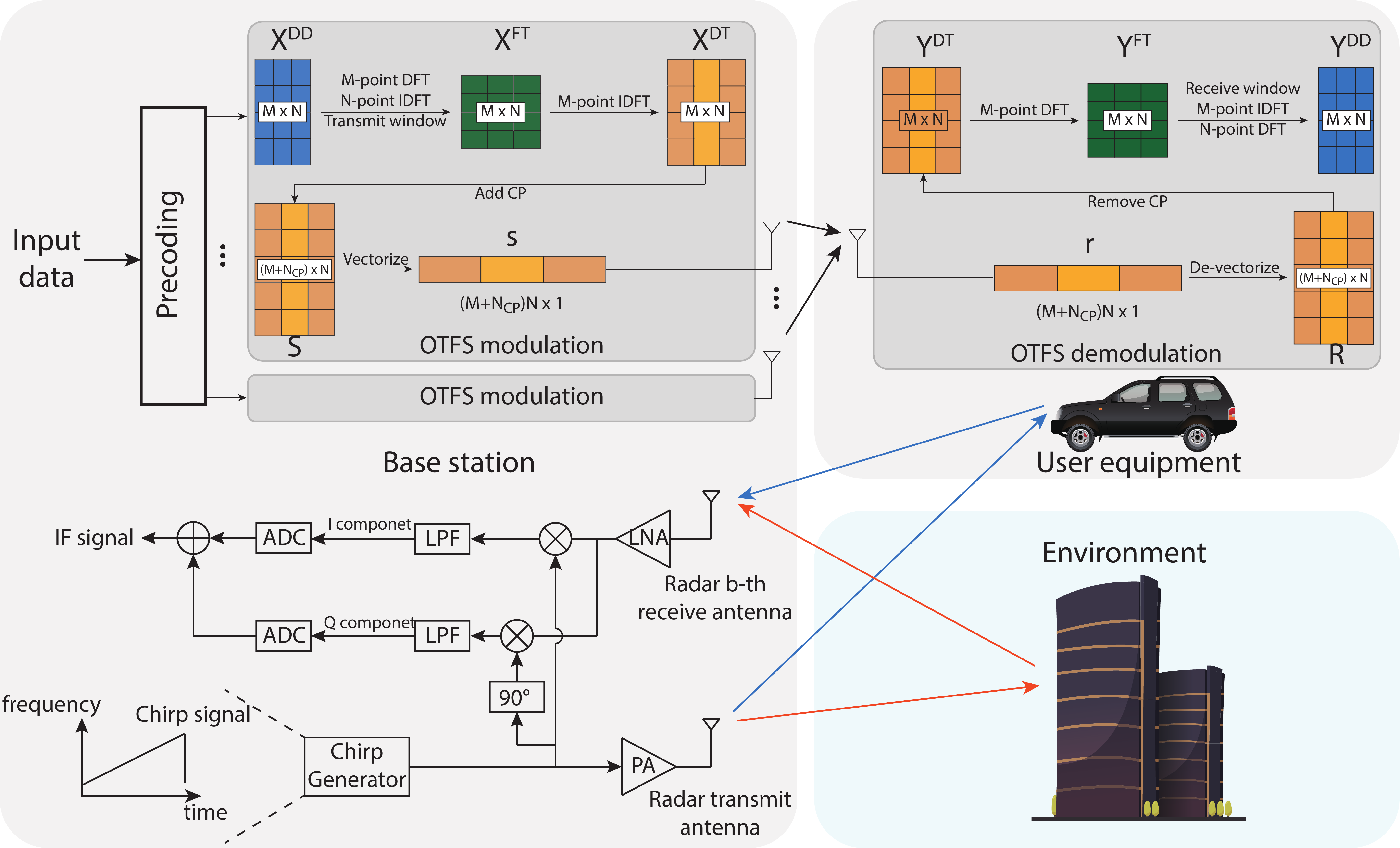}
		\caption{An illustration of the adopted signal models for both the communication and radar systems. }
		\label{fig:sigmodel}
	\end{figure}

	\subsubsection{OTFS Demodulation}
	At the user, the discrete-time time-domain basedband receive signal can be denoted by $\br\in\bbR^{(M+N_{\rm{CP}})N \times 1}$. In OTFS demodulation, the $\br$ is first rearranged into a two-dimensional signal $\bR$, which is given by
	\begin{align} \label{eq:time-domain receive}
	\bR = \mathrm{invec}(\br),
	\end{align}
	where $\mathrm{invec}(\cdot)$ denotes the inverse operation of $\mathrm{vec}(\cdot)$, \textit{i.e.}, \mbox{$\bA=\mathrm{invec} (\mathrm{vec}(\bA))$}. After that, the delay-time domain receive signal $\bY^{\rm{DT}}$ can be obtained by
	\begin{align}
	\bY^{\rm{DT}}= \bR_{\rm{CP}} {\bR},
	\end{align}
	where $\bR_{\rm{CP}} \in \bbR^{M\times (M+N_{\rm{CP}})}$ is the CP removal matrix. The time-frequency domain receive signal can then be written as
	\begin{align}
	\bY^{\rm{FT}} = \bW^{\rm{rx}} \odot \left( \bF_M \bY^{\rm{DT}}\right),
	\end{align}
	where $\bW^{\rm{rx}}$ is a windowing matrix, and we also adopt the all-one matrix for $\bW^{\rm{rx}}$. Finally, given $\bY^{\rm{FT}}$, the delay-Doppler receive signal ${\bY}^{\rm{DD}}$ is obtained by
	\begin{align}\label{eq:DD receive}
	{\bY}^{\rm{DD}} = \left({\bF}_{M}^{\rm H}{\bY}^{\rm{FT}} {\bF}_{N}\right).
	\end{align}
	\subsection{OTFS Channel Model}
	We consider a wide-band time-varying channel model incorporating $L$ propagation paths. Let $\alpha_i$, $\tau_i$, $\nu_i$, and $\psi_i$ denote the complex gain, the delay, the Doppler frequency shift, and the angle of departure (AoD) associated with the $i$-th ($i \in [1, \hdots,L$]) path, respectively. Let $M\Delta f$ denote the bandwidth of the OTFS system, and $NT$ denote the time duration of one OTFS frame. The delay tap index $m_i$ and the Doppler tap index $n_i$ corresponding to the $i$-th path can then be written as
	\begin{subequations}\label{eq:channel_tap}
		\begin{equation}
		m_i=round(M\Delta f \tau_i)
		\end{equation}
		\begin{equation}
		n_i=round(NT\nu_i),
		\end{equation}
	\end{subequations}
	where $round(\cdot)$ denotes the rounding operation. 
	Note that, for simplicity, this paper only considers the integer delay and Doppler cases, \textit{i.e.}, $m_i \in \bbZ$ and $n_i \in \bbZ$. The discrete delay-time baseband channel of the $a$-th \mbox{($a \in [1, \hdots,A]$)} transmit antenna can then be written as
	\begin{align}
	h[m, q, a]=\sum_{i=1}^{L} \alpha_i z^{n_i(q-m)} e^{-j 2 \pi\frac{d}{\lambda} (a-1) \psi_i} \mathrm{sinc}(m-m_i),
	\end{align}
	where $m\in [0, \hdots, M+N_{\rm{CP}}-1]$ denotes the index of the delay tap, $q \in [0, \hdots, N(M+N_{\rm{CP}})-1]$ denotes the index of the time tap, and \mbox{$z = \frac{j2 \pi}{N(M+N_{\rm{CP}})}$}.
	With this channel model, the receive signal at the user can be written as
	\begin{align} \label{eq:MIMO signal model}
	r[q] = \sum_{a=1}^{A} \sum_{m=0}^{M-1} h[m, q, n] s_a[q-m] + v[q]
	\end{align}
	where $r[q]$ denotes the $q$-th element in $\br\in\bbR^{(M+N_{\rm{CP}})N \times 1}$, and $s_a[q-m]$ denotes the $(q-m)$-th symbol transmitted by the $a$-th transmit antenna. The noise at the $q$-th time tap is denoted by $v[q]$.
	\par
	\textbf{OTFS Delay-Doppler Domain Channel Effect:}
	Let $Y_{m, n}^{\mathrm{DD}}$ denote the element at the $m$-th row and the $n$-th column in $\bY^{\mathrm{DD}}$, and $X_{m^\prime, n^\prime, a}^{\mathrm{DD}}$ denote the element at the $m^\prime$-th row and the $n^\prime$-th column in $\bX^{\mathrm{DD}}$ transmitted by the $a$-th antenna. Then, the input-output relation between the delay-Doppler domain signal $\bX^{\mathrm{DD}}$ in \eqref{eq:DD transmit} and the $\bY^{\mathrm{DD}}$ in \eqref{eq:DD receive} can be written as
	\begin{align} \label{eq:MIMO DD signal model}
	Y_{m, n}^{\mathrm{DD}} &= \sum_{a=1}^{A} \sum_{m^{\prime}=0}^{M-1} \sum_{n^{\prime}=-N / 2}^{N / 2-1} X_{m^{\prime}, n^{\prime}, a}^{\mathrm{DD}} H_{m-m^{\prime}, n-n^{\prime}, a}^{\mathrm{DD}}\, z^{l \left(n-n^{\prime}\right)_N} \nonumber\\
	&+V_{m, n}^{\mathrm{DD}},
	\end{align}
	where $V_{m, n}^{\mathrm{DD}}$ is the noise in the delay-Doppler domain. Note that $(n)_N$ denotes the modulo operation. The $H_{m, n, a}^{\mathrm{DD}}$ in \eqref{eq:MIMO DD signal model} is the delay-Doppler domain channel coefficient corresponding to the $m$-th delay tap, $n$-th Doppler tap, and the $a$-th transmit antenna, which satisfies
	\begin{align} \label{eq:DD channel}
	H_{m, n, a}^{\mathrm{DD}}=\sum_{i=1}^{L} \alpha_i \delta\left(m_i-(m)_{M}, n_i - (n)_N\right) e^{-j 2 \pi\frac{d}{\lambda} (a-1) \psi_i}.
	\end{align}
	\subsection{Radar Signal Model}
	In our system model, the BS is equipped with an FMCW radar. For simplicity, we assume that the radar has a single transmit antenna and $B$ receive antennas. However, our radar signal model and processing can be generalized to MIMO radars. Since the sensing targets of our interest are usually located away from the radar, we can assume that they are in the far-field region of the radar.
	\par
	The function of this radar is to obtain sensing information about the surrounding environment. In particular, the FMCW radar first emits chirp signals into the surrounding environment. These chirp signals interact with the surrounding objects and are reflected/scattered back to the radar. The received chirp signals are then processed to extract the sensing information. Mathematically, a transmitted radar chirp signal can be expressed as \cite{asuzu2018more}
	\begin{equation}
	s_{\text {chirp }}(t)= \begin{cases}\cos \left( 2 \pi f_0 t+\pi S t^{2} \right) & \text { if } 0 \leq t \leq T_{c} \\ 0 & \text { otherwise }\end{cases},
	\end{equation}
	where $f_0$, $S$, and $T_{c}$ represent the start frequency, the slope, and the duration of the chirp signal, respectively. Note that the frequency of the chirp signal linearly increases from $f_0$ to $f_0 + ST_c$ during the transmission. The effective bandwidth of the chirp signal is given by $B_{\mathrm{w}}=ST_c$.
	\par
	To be able to obtain the Doppler/velocity information of the surrounding objects, the radar typically transmits $N_{\text{loop}}$ identical chirp signals. These identical chirp signals form a radar frame. The transmitted signal in one radar frame can then be written as
	\begin{equation}
	s_{\text{frame}}(t) = \sum_{n=0}^{N_{\text{loop}}-1}  s_{\text {chirp}} \left(t-nT_{p}\right), \quad 0 \leq t \leq T_{f},
	\end{equation}
	where $T_{p}$ denotes the chirp repetition time, and $T_{f}$ denotes the radar frame duration. Note that $T_{c} \leq T_{p}$ is satisfied so that the chirp signals are non-overlapping.
	After the transmitted signal is reflected/scattered back and captured by the receive antennas, the received signal at each antenna is mixed with the transmitted signal using a quadrature mixer. The outputs of the quadrature mixer are the in-phase and quadrature components. The in-phase and quadrature components are then passed through a low-pass filter to obtain the so-called intermediate frequency (IF) signal. Assuming $W$ ideal point reflectors to be the sensing targets, the IF signal corresponding to a single chirp at the $b$-th ($b \in [1, \hdots, B]$) antenna can be written as
	\begin{align} 
	&r_{\text{chirp}}^b(t) \nonumber \\
	\approx& \sum_{w=1}^W \beta_w \exp(j2\pi f_0 \tau_w - j\pi S\tau_w^2)\exp(j2\pi S \tau_w t) e^{-j 2 \pi\frac{d_{\mathrm{r}}}{\lambda_{\mathrm{r}}} (b-1)} \label{eq:receive_chirp1} \\
	\approx& \sum_{w=1}^W \beta_w \exp(j2\pi f_0 \tau_w)\exp(j2\pi S \tau_w t) e^{-j 2 \pi\frac{d_{\mathrm{r}}}{\lambda_{\mathrm{r}}} (b-1)}\label{eq:receive_chirp},
	\end{align}
	where $d_{\mathrm{r}}$ and $\lambda_{\mathrm{r}}$ are the wavelength and the antenna spacing of the radar.
	The $\beta$ is the complex gain that depends on the radar cross section (RCS), the transmit power, and the path-loss. $\tau_w=\frac{2D_w}{c}$ is the round-trip propagation delay with $D_w$ denoting the propagation distance between the radar and \mbox{$w$-th} ideal point reflector. $c$ represents the speed of light. In \eqref{eq:receive_chirp1}, we neglect the receive signals that have interacted with multiple sensing targets since they have smaller power. The approximation in \eqref{eq:receive_chirp} holds when \mbox{$S\tau \ll f_0$}.
	\par
	The receive signal at each antenna $r^b_{\text{chirp}}(t)$ is then sampled by ADCs with the sampling rate of $f_s$. Let $N_s$ denote the number of complex ADC samples for each chirp. Note that each receive antenna has an independent receive chain including a quadrature mixer, low-pass filter, and ADCs. Finally, the ADC samples corresponding to $N_{\text{loop}}$ chirps and $B$ receive antennas are collected to form a radar data frame denoted by $\bX \in \bbC^{N_s\times N_{\text{loop}} \times B}$.
	\section{MIMO-OTFS Downlink Channel Estimation}\label{sec:MIMO-OTFS Downlink Channel Estimation}
	\sref{sec:OTFS signal model} presented the OTFS signal model. Particularly, the delay-Doppler channel is given by \eqref{eq:DD channel}, and the delay-Doppler domain input-output relation is given by \eqref{eq:MIMO DD signal model}. In this section, we first present the adopted delay-Doppler domain  OTFS frame structure. After that, we introduce the formulation of the delay-Doppler domain channel estimation problem.
	\subsection{OTFS Frame Structure}
\begin{figure}[t]
		\centering
		\includegraphics[width=0.85\linewidth]{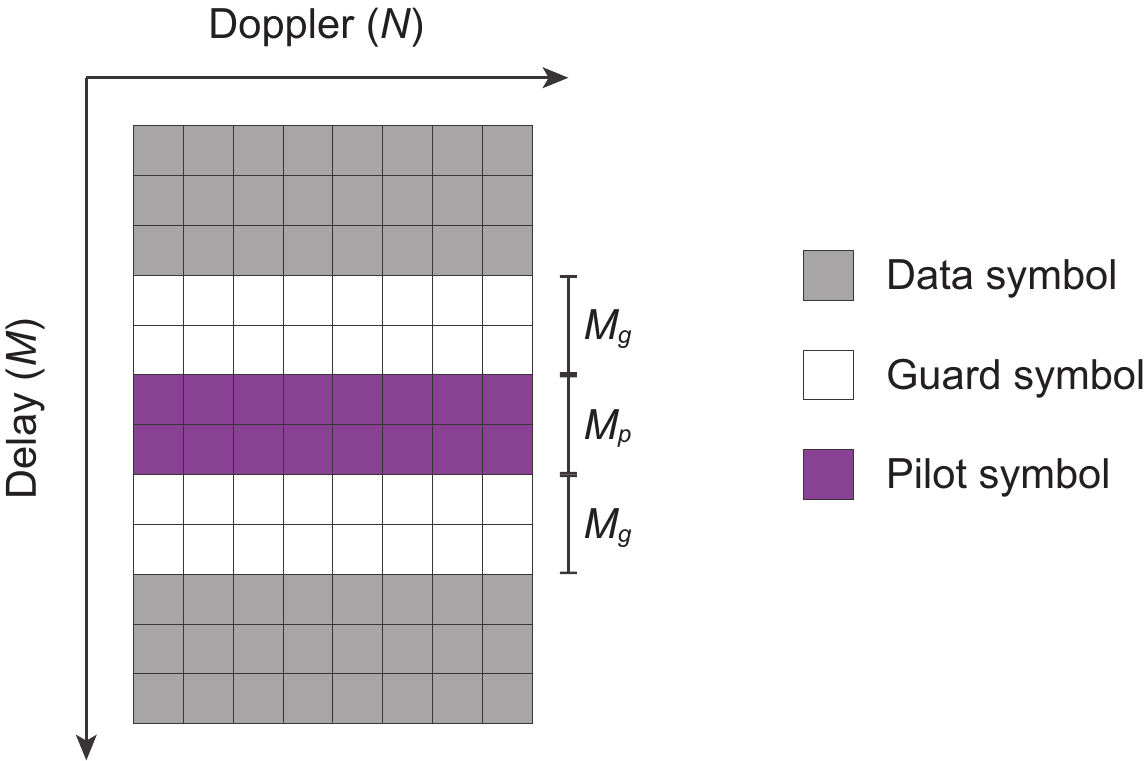}
		\caption{This figure shows the adopted structure of the delay-Doppler frame with the data, pilot, and guard symbols.}
		\label{fig:pilot}
	\end{figure}

	Fig. \ref{fig:pilot} shows the adopted pilot OTFS frame structure. The pilot and data symbols co-exist in the same OTFS frame. For simplicity, the data and pilot symbols are assumed to span across the entire Doppler dimension. The zero-power guard symbols are used to separate the pilot and data symbols along the delay dimension. The lengths of the pilot and guard symbols along the delay dimensions are $M_p$ and $M_g$ respectively. With $M_{\rm{max}}$ denoting the maximum delay spread, the length of the guard symbols along the delay dimension is assumed to satisfy $M_{\rm{g}} \geq M_{\rm{max}}$. This way, the guard symbols can guarantee that the data symbols do not interfere with the pilot symbols. Note that all the transmit antennas transmit pilot signals on the pilot resources simultaneously. The pilot overhead of the adopted OTFS frame structure can be described by the pilot overhead ratio $\eta = \frac{M_p}{M}$. A good pilot design/channel estimation strategy will attempt to minimize the pilot overhead ratio.
	
	\begin{figure*}[t]
	\centering
	\subfigure[LoS and NLoS communication paths]{\label{fig:commradarchannel1}\includegraphics[width=0.32\linewidth]{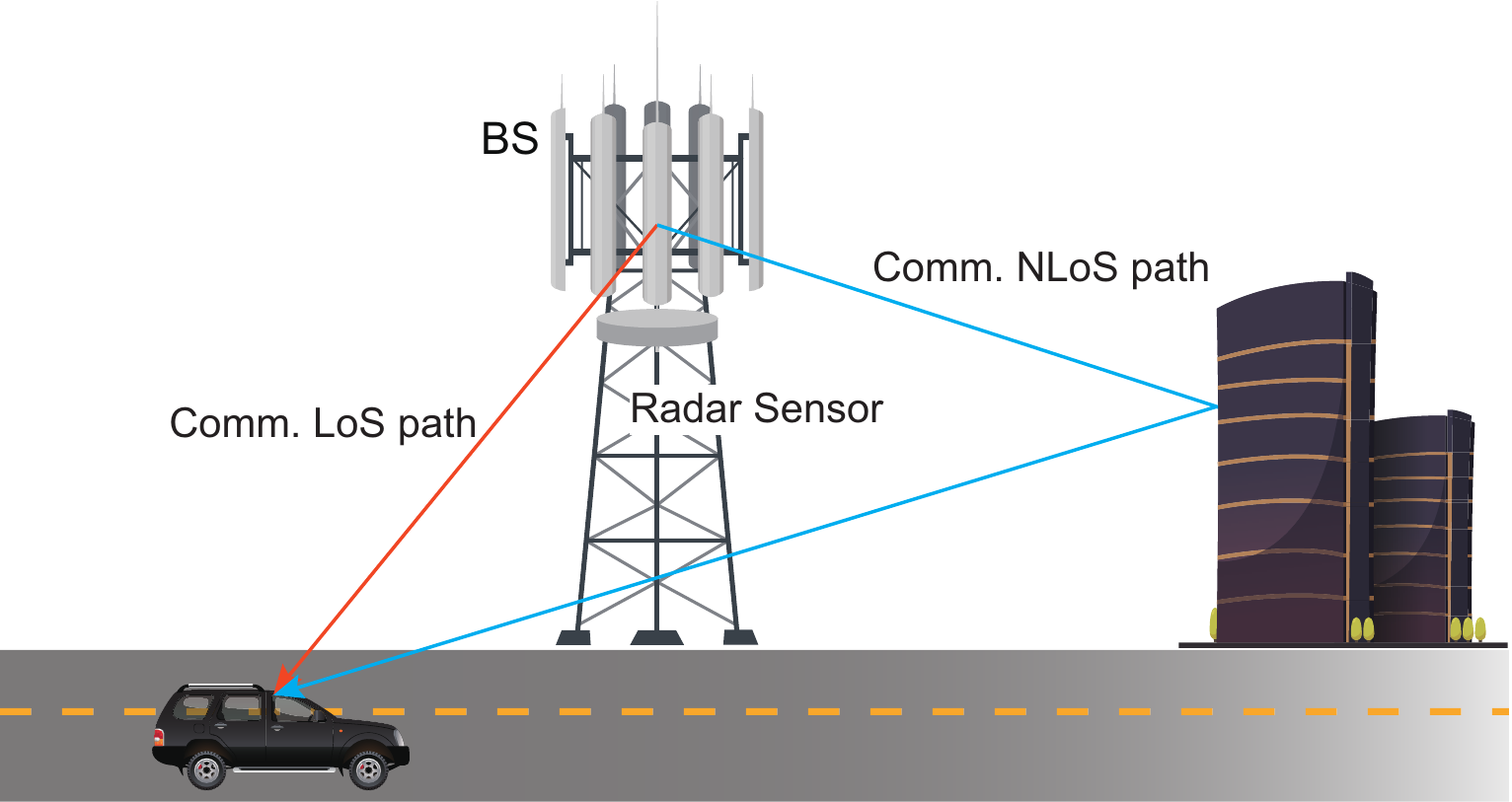}}
	\subfigure[Backscattering radar paths]{\label{fig:commradarchannel2}\includegraphics[width=0.32\linewidth]{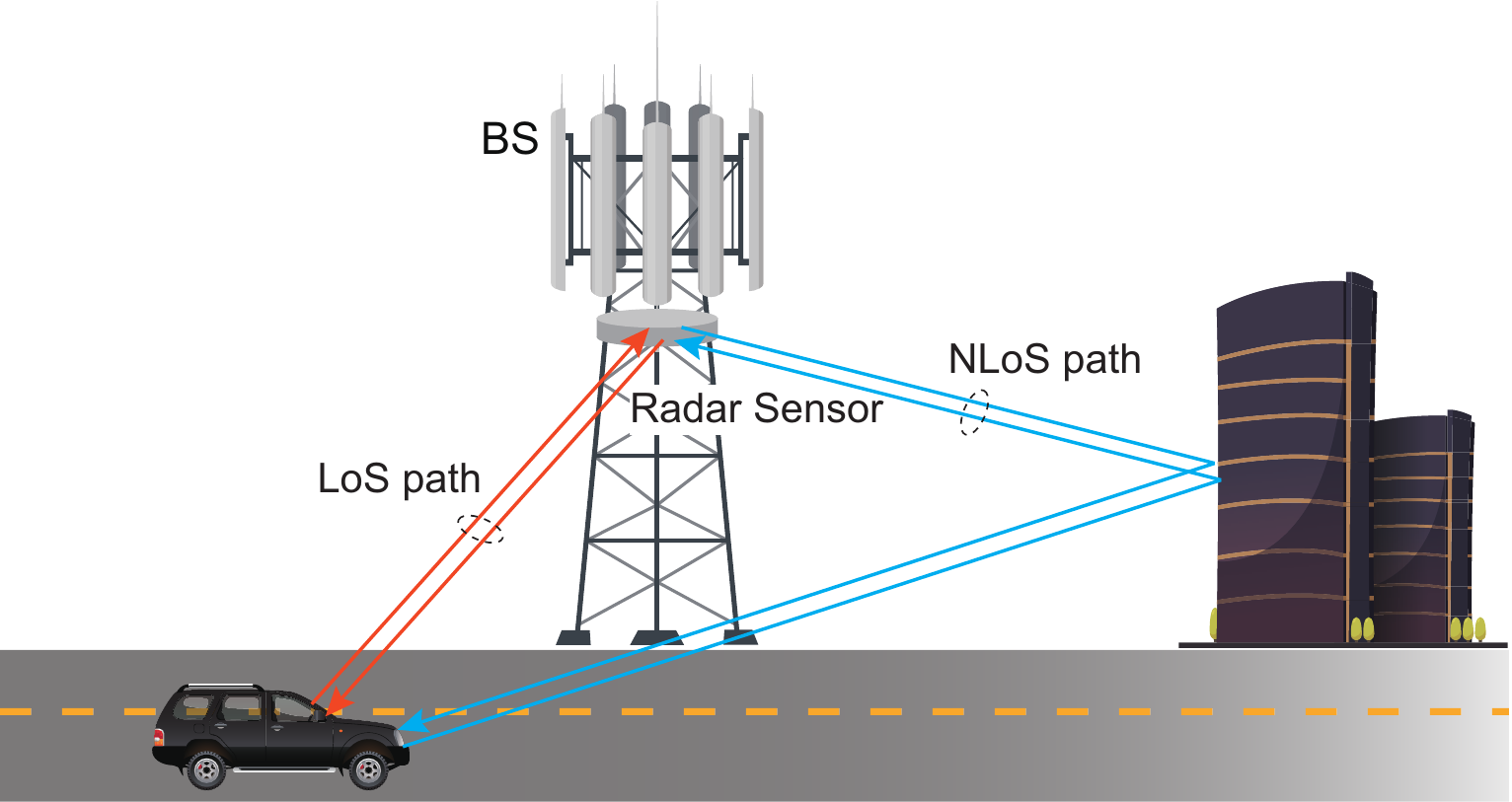}}
	\subfigure[Radar transmit and receive paths could be different]{\label{fig:commradarchannel4}\includegraphics[width=0.32\linewidth]{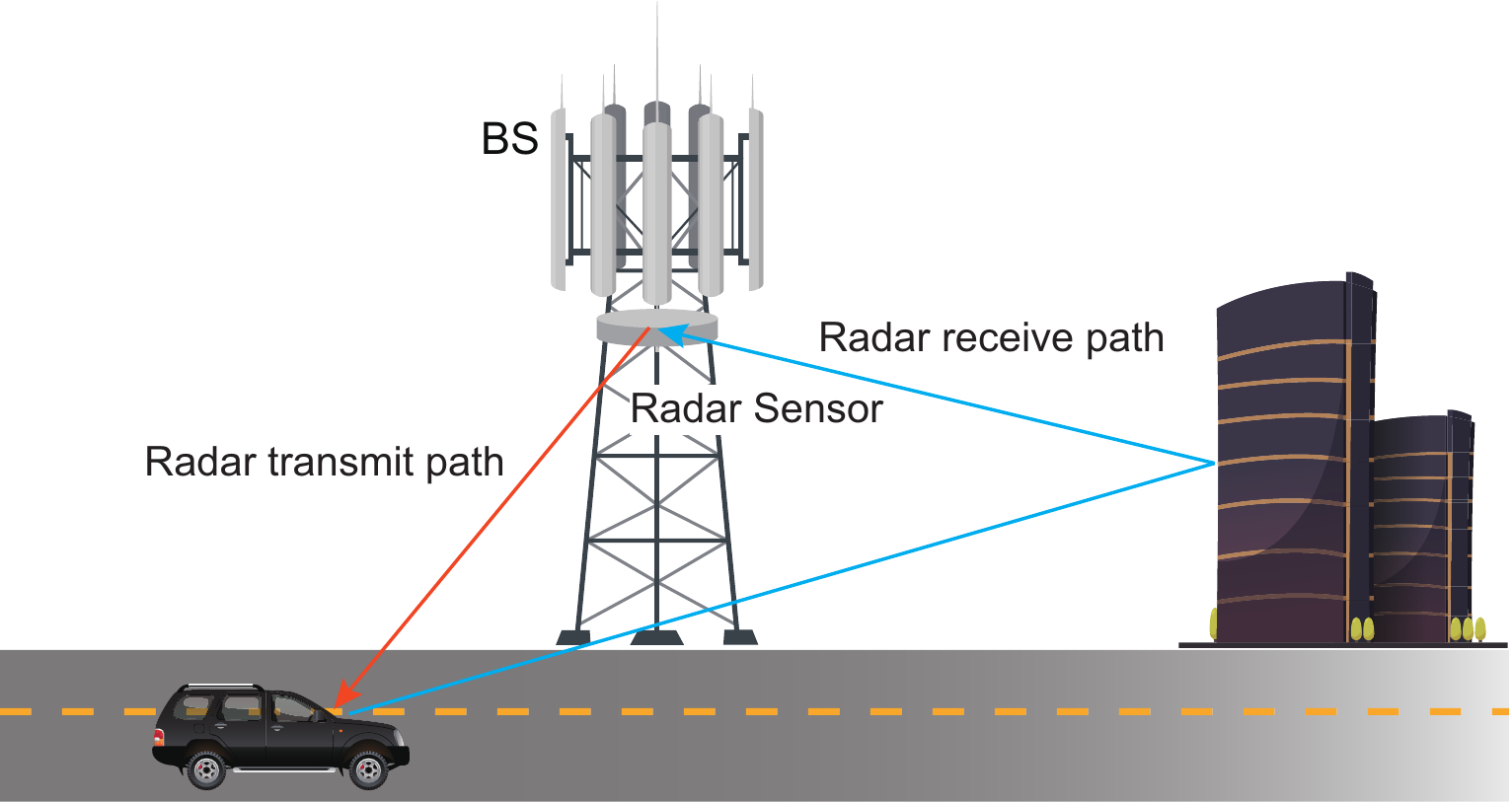}}
	\caption{This figure illustrates an example of the propagation paths for the radar and the downlink  communication systems. The transmitted radar signals could be backscattered to the receiver through the same transmit paths or reflected and received from different directions.}
	\label{fig:commradarchannel}
\end{figure*}

	\subsection{MIMO-OTFS Downlink Channel Estimation: A Sparse Recovery  Formulation}
	Let $x_{m,n,a}$ denote the training pilots in the delay-Doppler domain transmitted by the $a$-th antenna, where \mbox{$m\in[0,\hdots,M_p-1]$} is the pilot index along the delay dimension and \mbox{$n\in[-\frac{N}{2},\hdots,\frac{N}{2}-1]$} is the pilot index along the Doppler dimension. Derived from \eqref{eq:MIMO DD signal model}, the received signal $y_{m,n}$ can be written as
	\begin{align}\label{eq:tmpp}
	&y_{m, n} \nonumber \\
	=&\sum_{a=1}^{A} \sum_{m^{\prime}=0}^{M_{\rm{g}}-1} \sum_{n^{\prime}=-N / 2}^{N / 2-1} z^{n \left(m-m^{\prime}\right)_M} H_{m^{\prime}, n^{\prime},a}^{\mathrm{DD}}\, x_{(m-m^{\prime})_M, (n-n^{\prime})_N, a}\nonumber\\
	&+v_{m, n}.
	\end{align}
	Next, the pilot symbols $x_{m,n,a}$, delay-Doppler channel coefficients $H_{m, n, a}^{\mathrm{DD}}$, and received pilot $y_{m,n}$ symbols are rearranged, and \eqref{eq:tmpp} is rewritten as a matrix-vector multiplication to form a sparse problem.
	\par
	Let $\bZ \in \bbC^{M_p N \times A M_{\rm{g}}N}$ with the $(mN+n+N/2+1, A (m^\prime N+n^\prime+N/2)+a)$-th element being $z^{n \left(m-m^{\prime}\right)_M}$. Let $\bP \in \bbC^{M_p N \times A M_{\rm{g}}N}$ with the $(mN+n+N/2+1, A(m^\prime N+n^\prime+N/2)+a)$-th element being $x_{(m-m^{\prime})_M, (n-n^{\prime})_N, a}$. Then \eqref{eq:tmpp} can be re-written as
	\begin{align}\label{eq:tmp}
	\by = \left(\bZ \odot \bP \right) \bh + \bv,
	\end{align}
	where $\by \in \bbC^{M_p N \times1}$ with the $(m N+n+N/2+1)$-th element being $y_{m,n}$, and $\bh \in \bbC^{AM_{\rm{g}} N \times1}$ with the $(A(m^\prime N+n^\prime+N/2)+a)$-th element being $H_{m^{\prime}, n^{\prime},a}^{\mathrm{DD}}$.
	\par
	Let ${\bf \Psi} = \bZ \odot \bP$, then following \cite{shen2019channel}, the sparse  formulation is given by
	\begin{align}\label{eq:p1}
	\by = {\bf \Psi} \bh + \bv.
	\end{align}
	Note that each element in $\bh$ corresponds to a delay and Doppler tap in the delay-Doppler domain, and $\bh$ is a sparse vector due to the sparsity of the delay-Doppler channel.
	According to \cite{3GPPchannel}, the number of dominant propagation path is limited (\textit{e.g.} 6 paths). Therefore, the delay-Doppler channel is sparse in the delay dimension. The delay-Doppler channel is also sparse in the Doppler dimension since the Doppler frequency of a path is usually much smaller than the system bandwidth, and only the near-zero Doppler taps have relatively high power. Moreover, the transmit antenna domain can be further converted to the virtual angle domain to increase the sparsity of the $\bh$. Let $\bA = \bI_{M_{\rm{g}} N}\otimes \bF_{A}$, with $\otimes$ denoting the Kronecker product, and $\bI$ denoting the identity matrix. The angle domain sparse formulation is then given by
	\begin{align}\label{eq:p2}
	\by = \tilde{\bf \Psi} \tilde{\bh} + \bv,
	\end{align}
	where $\tilde{\bf \Psi} = \bf \Psi \bA^{\rm H}$ and $\tilde{\bh} = \bA \bh$.
	Utilizing the sparsity of the delay-Doppler domain channel \cite{shen2019channel}, the MIMO-OTFS channel estimation can be achieved by solving \eqref{eq:p1} or \eqref{eq:p2} using conventional CS recovery algorithms such as basis pursuit \cite{BP} and matching pursuit \cite{MP}. Next, we discuss the key idea of utilizing sensing to guide the OTFS-based MIMO channel estimation.

	\section{Sensing Aided Delay-Doppler Channel Estimation}\label{sec:main}
	The convergence of communication, sensing, and localization is considered one of the key features in 6G and beyond \cite{rappaport2019wireless}. The sensing and localization capabilities may not just support new interesting applications such as AR/VR and autonomous driving, but also provide rich information and awareness about the surrounding environment to aid the communication systems. Furthermore, as will be explained in \sref{sec:key_idea}, this sensing information can be particularly meaningful and beneficial for delay-Doppler communication systems. To that end, we propose to utilize the sensing capability at the BS to aid the MIMO-OTFS channel estimation problem.
	
	\begin{figure*}[t]
		\centering
		\includegraphics[width=0.9\linewidth]{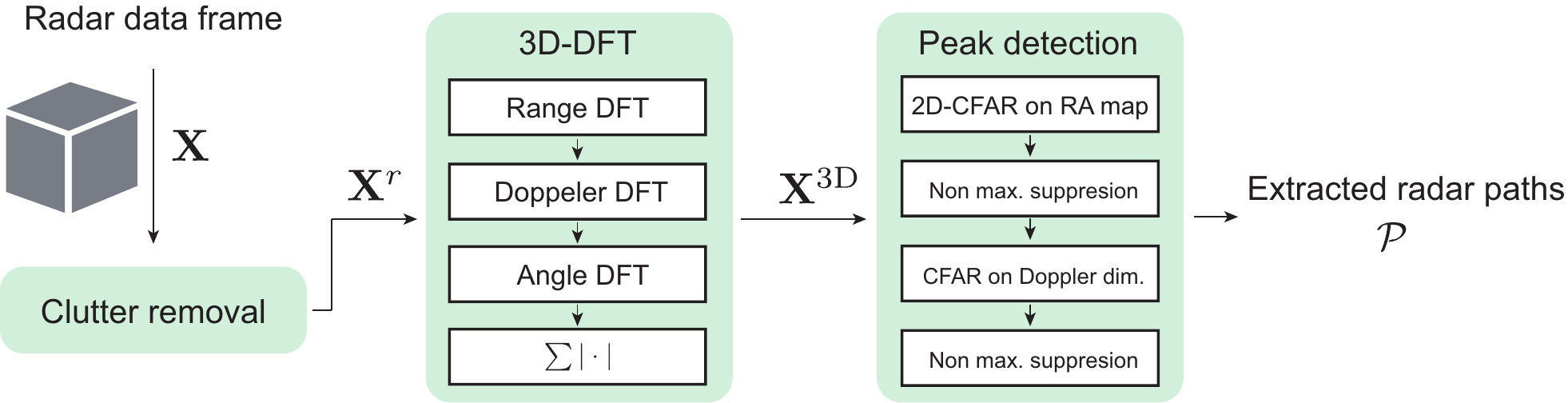}
		\caption{This figure presents the adopted radar processing. The radar data frame is processed through clutter removal, 3D-DFT, and peak detection to extract the radar propagation paths.}
		\label{fig:radar_processing}
	\end{figure*}

	In this section, we first briefly introduce the key idea of sensing-aided delay-Doppler communications. After that, we discuss the relationship between communication and radar channels. Then, we explain the adopted radar processing to extract sensing information. Last, we elaborate on the proposed sensing-aided channel estimation.
	\subsection{Key Idea: Sensing Aided Delay-Doppler Communications}\label{sec:key_idea}
	The delay-Doppler domain channel has a close and direct relation to the direction/position of the UEs and the geometry of the surrounding environment. In particular, as shown in \eqref{eq:DD channel}, each tap of the delay-Doppler domain channel corresponds to an existing propagation path of a certain delay and a certain Doppler frequency shift. This motivates utilizing the sensing capability to obtain prior information about the communication channel and improve delay-Doppler communications. For instance, using the sensing capability, the BS can obtain/estimate the relative position and velocity of the UE, and also the positions and shapes of the reflecting/scattering objects in the surrounding environment. \textbf{With this sensing information, the BS can infer the potential propagation path parameters: the delay, the Doppler velocity, and the AoD/AoA.} This prior knowledge of the propagation paths can help the delay-Doppler communications in several ways: (i) guiding or even bypassing channel estimation, (ii) improving channel feedback, and (iii) enabling proactive resource allocation.
	\par
	In this paper, we are particularly interested in exploiting radar at the BS to obtain sensing information. Compared with other sensory options, radars have the following advantage. (i) Recently, joint communication and radar systems have gained increasing interest \cite{liu2020joint, feng2020joint}. Being able to share the hardware and software resources with the communication systems can make the radar a more available and low-cost sensing solution. (ii) Since the radar sensing signals are also transmitted through the wireless channel, the sensing information extracted from radars can have a closer and more straightforward relation to the wireless communication channels. (iii) Radar sensing can potentially obtain NLoS sensing information, which may not be available using other sensors.
	\par
	Next, we discuss the relationship between the communication channel and the radar channel.
	\subsection{Relation Between Communication and Radar Channels}

	In \figref{fig:commradarchannel}, we present a scenario incorporating a BS, a UE, and a static reflector. \figref{fig:commradarchannel1} shows the  line-of-sight (LoS) path and the non-line-of-sight (NLoS) paths of communication and radar channels. In \figref{fig:commradarchannel2}, two sets of radar propagation paths resulting from backscattering are presented. In this backscattering case, the radar propagation paths are closely related to the communication propagation paths shown by \figref{fig:commradarchannel1} in terms of the propagation delay, the Doppler velocity, and the AoD/AoA. Note that, the radar signal is likely to be reflected by other components of the UE instead of the UE antenna. Therefore, the communication paths in \figref{fig:commradarchannel2} does not align perfectly with the radar paths in \figref{fig:commradarchannel1}. However, when the distance between the BS and the UE is much greater than the size of the UE, the relation between the communication and radar paths can be approximated as follows.
	\begin{itemize}
		\item The \textbf{propagation delays and Doppler velocities} of the radar paths are approximately twice of those of the corresponding communication paths.
		\item The \textbf{AoDs/AoAs} of radar paths are approximately the same as the AoDs of the corresponding communication paths.
	\end{itemize}
	\par
	Apart from the backscattering cases shown in \figref{fig:commradarchannel2}, the radar transmit and receive propagation paths may be completely different. For instance, as shown in both \figref{fig:commradarchannel4}, the radar transmit and receive signals propagate through the LoS and NLoS paths, respectively. As a result, the delay, the Doppler velocity, and the AoA/AoD of the radar propagation paths are not directly related to one communication path. Although the radar propagation paths shown in \figref{fig:commradarchannel4} can lead to interference in the radar receive signals, they can be identified and filtered by exploiting the fact that the AoD and AoA are different. This can be potentially achieved when the radar has the transmit beamforming capability. However, this case shown in \figref{fig:commradarchannel4} is not handled by this paper.
	\subsection{Radar Processing}
	From the captured radar data frame $\bX$, we aim to extract the propagation delay, Doppler velocity, and angle of arrival (AoA) of the radar propagation paths corresponding to the UE. The proposed radar processing is illustrated in \figref{fig:radar_processing}.
	\par
	\textbf{Cutter removal}: Since we are interested in high-mobility scenarios, the clutter removal is first applied to the radar data frame $\bX$ to remove the radar signals corresponding to static objects. The clutter removal is mathematically given by
	\begin{equation}
	X^r_{m,n,b} = X_{m,n,b} - \frac{1}{N_{\text{loop}}}\sum_{n=1}^{N_{\text{loop}}} X_{m,n,b},
	\end{equation}
	where $\bX^r$ denotes the radar data frame after clutter removal. $X_{m,n,b}$ and $X^r_{m,n,b}$ index the elements from $\bX$ and $\bX^r$ according to the indices $m$, $n$, and $b$. 
	\par
	\textbf{3D-DFT:} After the clutter removal, we sequentially apply three DFTs on $\bX^r$ to extract the range, Doppler, and angle information corresponding to the moving objects in the radar data frame.
	\begin{itemize}
		\item Range DFT: we first apply the DFT on the ADC samples dimension of $\bX^r$. This converts the chirp signal into the frequency domain. As can be observed from \eqref{eq:receive_chirp}, the frequency of the received chirp signal is proportional to the propagation delay.
		\item Doppler DFT: After the range DFT, we apply the Doppler DFT along the second dimension of the $\bX^r$. The Doppler DFT obtains the phase shift across the consecutive chirp signals. From these phase shifts, the Doppler velocity of the moving objects can be extracted.
		\item Angle DFT: The angle DFT operation is performed on the radar virtual antenna dimension, which extracts the angular information of the moving objects. Note that zero-paddings can be applied before the angle DFT for more accurate angular estimation.
	\end{itemize}
	With $\text{DFT}_{\text{3D}}$ denoting the three DFT operations on the range, Doppler, and angle dimensions, the radar 3D-heatmap can be obtained by
	\begin{align}
	\bX^{\text{3D}} = |\text{DFT}_{\text{3D}} \left( \bX^r \right)|,
	\end{align}
	where the absolute operation is applied element-wise.

	\textbf{Peak detection}: 
	To detect the peaks in the 3D-heatmap $\bX^{\text{3D}}$, we employ the constant rate false alarm rate (CFAR)	algorithm \cite{rohling1983radar}. Since the 3D-CFAR is computationally expensive, we first apply a 2D-CFAR on the range-angle heatmap $\bX^{\text{RA}}$. The range-angle heatmap can be obtained by averaging the Doppler dimension of $\bX^{\text{3D}}$ as shown by
	\begin{align}
	X^{\text{RA}}_{m,b} = \frac{1}{N_{\text{loop}}}\sum_{n=1}^{N_{\text{loop}}} X^{\text{3D}}_{m,n,b}.
	\end{align}
	After that, a non-maximum suppression is applied to the peaks detected by the 2D-CFAR to deal with the power leakage along the range and angle dimensions. Then we apply a 1D-CFAR on the Doppler dimensions of $\bX^{\text{3D}}$ according to the peaks detected on the range-angle heatmap. The non-maximum suppression is also employed after the 1D-CFAR.
	\par
	After the peak detection, the propagation delay, Doppler velocity, and AoA of each peak is extracted. Let $J$ denote the number of detected peaks from the radar peak detection, we can obtain $\cP = \{ (\tau^p_1, v^p_1, \theta^p_1), \hdots, (\tau^p_{J}, v^p_{J}, \theta^p_{J})\}$, where $\tau^p_{j}$, $v^p_{j}$, $\theta^p_{j}$ denote the propagation delay, Doppler velocity, and AoA of the $j$-th \mbox{$(j=[1,\hdots, J])$} peak, respectively.

	\subsection{Sparse Recovery with Radar Sensing Information}
	To utilize the sensing information obtained from the radar to MIMO-OTFS channel estimation, we first convert each peak in $\cP$ to the index of angle domain OTFS channel $\tilde{\bh}$ according to the propagation delay, Doppler velocity, and AoA of the peak. The propagation delay of the radar-detected propagation paths is normalized as follows:
	\begin{equation}
	\tilde{\tau}^p_j = \frac{\tau^p_j - \min_{j} \tau^p_j}{2}.
	\end{equation}
	The above normalization is based on that the shortest radar propagation path can be detected as a peak in $\cP$. This is a reasonable assumption since the shortest propagation path is likely to be one of the strongest paths.
	To convert the Doppler velocity of each detected peak to Doppler frequency, the Doppler velocity is multiplied by the carrier frequency of the communication system as shown by
		\begin{algorithm}[t]
		\caption{\label{tbl:proposed}OMP Algorithm using Radar Sensing}
		\begin{algorithmic}[1]
			\INPUT\, Received pilot signal $\by$
			
			sensing matrix $\tilde{\bf \Psi}$
			
			receive radar data frame $\bX$
			
			threshold $\rho$
			\OUTPUT recovered MIMO-OTFS channel $\tilde{\bh}$
			\STATE Obtain $\cS^r$ from $\bX$ using radar processing
			\STATE \textbf{Initialization} $S^{(0)} = \cS^r$, $\tilde{\bh}^{(0)}={\bf 0}$, $iter=1$,
			
			\quad\quad \quad \quad$\bb = \by - \tilde{\bf\Psi} \tilde{\bh}^{(0)}$
			\WHILE{$|\cS| < \rho \cdot |\cS_r|$}
			\STATE $\bz = \tilde{\bf\Psi}^{H} \bb$
			\STATE $\cS \leftarrow \cS \cup \arg\max_j z[j]^{(iter)}$
			\STATE $\tilde{\bh}^{(iter)}_{S} = \tilde{\bf \Psi}_{S}^{\dagger}\by$, $\tilde{\bh}^{(iter)}_{S^c} = 0$
			\STATE $\bb = \by - \tilde{\bf\Psi} \tilde{\bh}^{(iter)}$
			\STATE $iter = iter + 1$
			\ENDWHILE
		\end{algorithmic}
	\end{algorithm}

\begin{figure}[t]
	\centering
	\includegraphics[width=1\linewidth]{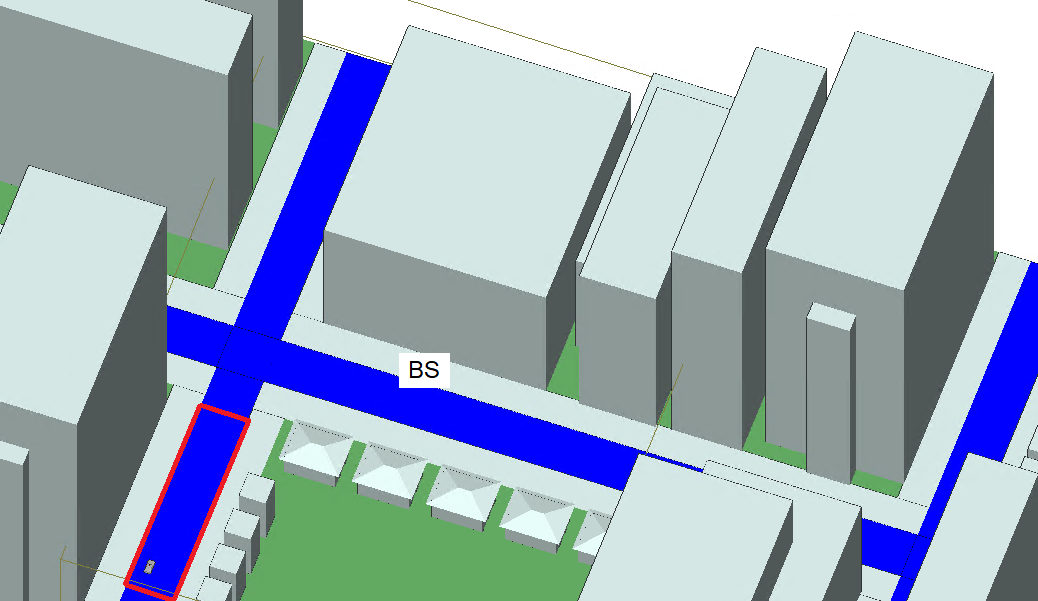}
	\caption{This figure shows the developed 3D ray-tracing scenario. The BS is deployed along the horizontal street, and the UE is randomly dropped (in different locations for the different realizations) within the area defined by the red rectangle.}
	\label{fig:layout}
\end{figure}

	\begin{equation}
	\tilde{\nu}^p_j = \frac{v^p_j}{2} \cdot f_c.
	\end{equation}
	The normalized propagation delay $\tilde{\tau}^p_j$ and Doppler frequency $\tilde{\nu}^p_j$ are converted to delay tap and Doppler tap indices $m^p_j$ and $n^p_j$ of the communication channel using
	\eqref{eq:channel_tap}, respectively. The AoA $\theta^p_j$ of the radar detected peaks are converted to the row indices $f_j$ of $\bF_A$ according to the beam steering angle of the row vectors in $\bF_A$. Mathematically, $f_j$ is given by
	\begin{align}
	&f_j = \nonumber \\
	&\underset{f_j}{\arg\max} \  \bigg |\bff_{f_j}   [1, e^{-j 2 \pi\frac{d}{\lambda} \cos\theta^p_j} , \hdots, e^{-j 2 \pi\frac{d}{\lambda} (A-1) \cos\theta^p_j}]^T \bigg|^2,
	\end{align}
	where $\bff_{f}$ denotes the $f$-th row of $\bF_{A}$. Finally, the $\cP$ is converted to set $\cS_r = \{t_1, \hdots, t_{J}\}$, where $t_j$ is the index of $\tilde{\bh}$ corresponding to the $j$-th detected peak in $\cP$. $t_j$ can be obtained by
	\begin{equation}
	t_j = f_j + (A-1) m^p_j +  (A-1) (M_g-1) ( n^p_j + \frac{N}{2}).
	\end{equation}
	After that, the $\cS_r$ is sent to the UE over the control channel.
	\par
	At the UE side, we propose to utilize the radar extracted sensing information $\cS_r$ to improve sparse channel recovery. The proposed sparse channel recovery approach is presented in Algorithm \ref{tbl:proposed}, which is a modification of the OMP algorithm \cite{tropp2007signal}. The OMP iteratively updates the estimated support of the sparse signal by calculating the strongest correlations between the residual signal and the sensing matrix. After the last iteration, the OMP applies the least square (LS) estimation on the elements in the sparse signal as indicated by the estimated support.
	Based on the OMP algorithm, we make two modifications. First, we use the $\cS_r$ to initialize the estimated support of the sparse signal. The delay, Doppler, and AoAs taps extracted from the radar sensing information are expected to be similar to those of the communication channels. Second, we set the maximum number of iterations and the size of the estimated support adaptively according to the number of peaks detected in $\cS_r$ as shown by
	\begin{equation}\label{eq:adapt}
	|\cS| \leq \rho \cdot |\cS_r|,
	\end{equation}
	where $\rho \geq 1$ is a hyper-parameter. The intuition is that, when the UE is in a more complicated environment with many potential reflectors, the communication channel is more likely to incorporate more (resolvable) propagation paths. Meanwhile, the radar should also tend to detect more peaks/paths.

	\section{Simulation Data Generation}\label{sec:Simulation Data}
	In this paper, we propose to utilize radar sensing capability at the BS to improve downlink channel estimation for MIMO-OTFS systems. Hence, realistic co-existing wireless communication and radar channel modeling is essential for our simulation. To that end, we generate wireless communication and radar channels based on accurate ray-tracing. \figref{fig:layout} shows the bird view of the adopted ray-tracing scenario. The ray-tracing scenario models a downtown area. It consists of the intersections of one horizontal street and two vertical streets, and various buildings. The BS is located on one side of the horizontal street and points to the other side of the street. The UE is randomly distributed in the area noted by the red box with a random velocity uniformly sampled from $[50, 90]\ m/s$. For communication and radar channels, we adopt the $6$GHz and $28$GHz frequency bands, respectively. The detailed simulation parameters are summarized in Table \ref{tbl:simu}. We sample 100 scenes with different UE locations and velocities. For each scene, we generate the communication and radar channel parameters. Specifically, based on ray-tracing, we simulate the parameters of each propagation path including the complex gain, the propagation delay, the Doppler frequency, the AoA, and the AoD. From these channel parameters, we can obtain the communication channel $H_{m,n,a}^{\text{DD}}$, and the radar data frame $\bX$.

	\begin{table}[!t]
		\caption{System Parameters for simulation}
		\label{tbl:simu}
		\centering
		\begin{tabular}{ll}
			\toprule
			- Wireless Communication Parameters- & \\
			Carrier frequency ($f_c)$&  $6\ GHz$\\
			Number of BS antennas ($A$) & $32$ \\
			Number of UE antennas & $1$ \\
			Number of subcarriers $(M)$ & 256\\
			Number of symbols $(N)$ & 14\\
			Cyclic prefix & $25\%$\\
			Subcarrier spacing & $15\ kHz$ \\
			UE velocity & $50 \sim 90\ m/s$\\
			\toprule
			- Radar Configuration - & \\
			Start frequency ($f_0$) & $28\ GHz$\\ 
			Slope ($S$) & $30\ MHz/\mu s$\\
			Sampling rate ($f_s$) & $30\ MHz$\\
			Effective bandwidth ($B_{\mathrm{w}}$) & $512\ MHz$\\
			Number of samples per chirp ($N_s$)  & $512$\\
			Number of loops ($N_{\text{loop}}$)& $256$\\
			Number of transmit antenna & $1$\\
			Number of receive antenna ($B$) & $16$\\
			\toprule
			- Radar Derived Parameters - & \\
			Max. unambiguous range & $149.6\ m$\\
			Range resolution & $0.29\ m$\\
			Max. unambiguous velocity & $\pm 121.47\ m/s$\\
			Velocity resolution & $0.96\ m/s$\\
			\bottomrule
		\end{tabular}
	\end{table}
	\begin{figure}[t]
	\centering
	\includegraphics[width=1\linewidth]{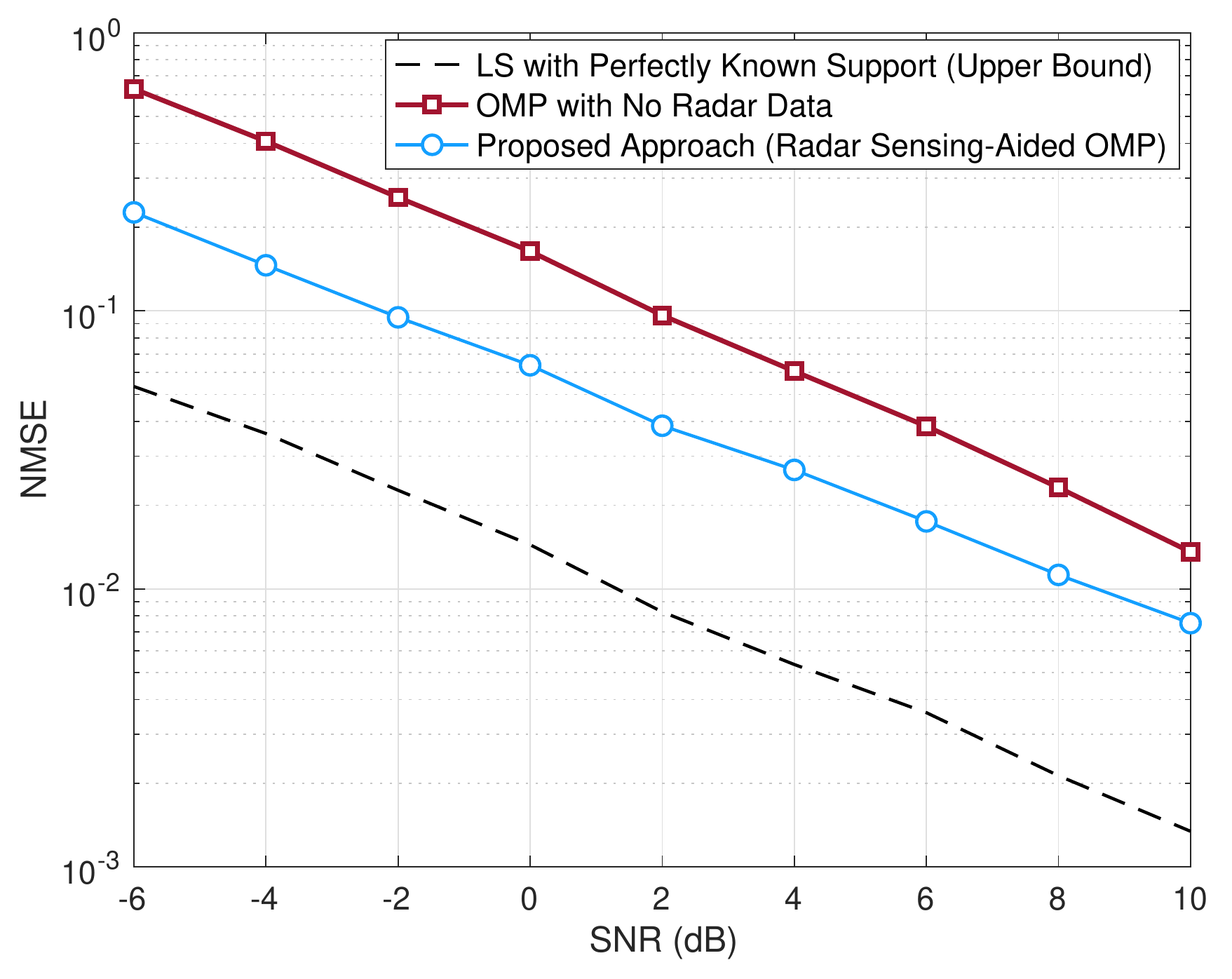}
	\caption{This figure shows the NMSE performance under different SNRs. The number of BS antennas is $32$ and the $\eta$ (pilot overhead ratio) is $20\%$.}
	\label{fig:snr}
\end{figure}

	\section{Simulation Results}\label{sec:Simulation Results}
	In this section, we investigate the performance of the proposed sensing-aided channel estimation compared with the conventional OMP in terms of the normalized mean square error (NMSE) and pilot overhead ratio.
	
	\subsection{Does Sensing Improve Channel Estimation Accuracy?}
	In Fig. \ref{fig:snr}, we show the NMSE performance comparison under  various SNR levels. The number of antennas is set to $32$, the pilot overhead ratio $\eta$ is $20\%$, \textit{i.e.}, $256 \times 14 \times  0.2\approx716$ symbols are allocated to the pilot signal. The genie-aided approach ``LS with perfectly known support" relies on perfect knowledge of the communication channel support to solve \eqref{eq:p1}. That is, the support of $\bh$ is assumed to be known, and LS is applied to recover the channel taps indicated by the support. Therefore, it acts as an approximated upper bound for the other solutions without perfect support knowledge. The ``OMP with no radar data" do not exploit any prior channel support information. Instead, it directly applies the OMP algorithm to solve \eqref{eq:p2}. The ``radar sensing-aided OMP" is the proposed approach  shown in Algorithm~\ref{tbl:proposed}.
	
	It can be observed that the NMSE performance of all three methods improve as the SNR increases. The proposed approach outperforms the conventional OMP (angle domain) in the SNR region shown in Fig. \ref{fig:snr}. In particular, \textbf{the proposed approach achieves similar NMSE performance with $\bf 4$dB lower SNR compared with the OMP (angle domain).} The performance gap between the proposed method and the OMP (angle domain) is slightly larger at the low SNR region. When the noise level is high, as it becomes difficult for the conventional OMP (angle domain) solution to estimate the correct dominant channel taps. However, the proposed method can leverage the channel support extracted from the radar signals, making it more robust even at low SNRs.
	
	\begin{figure}[t]
		\centering
		\includegraphics[width=1\linewidth]{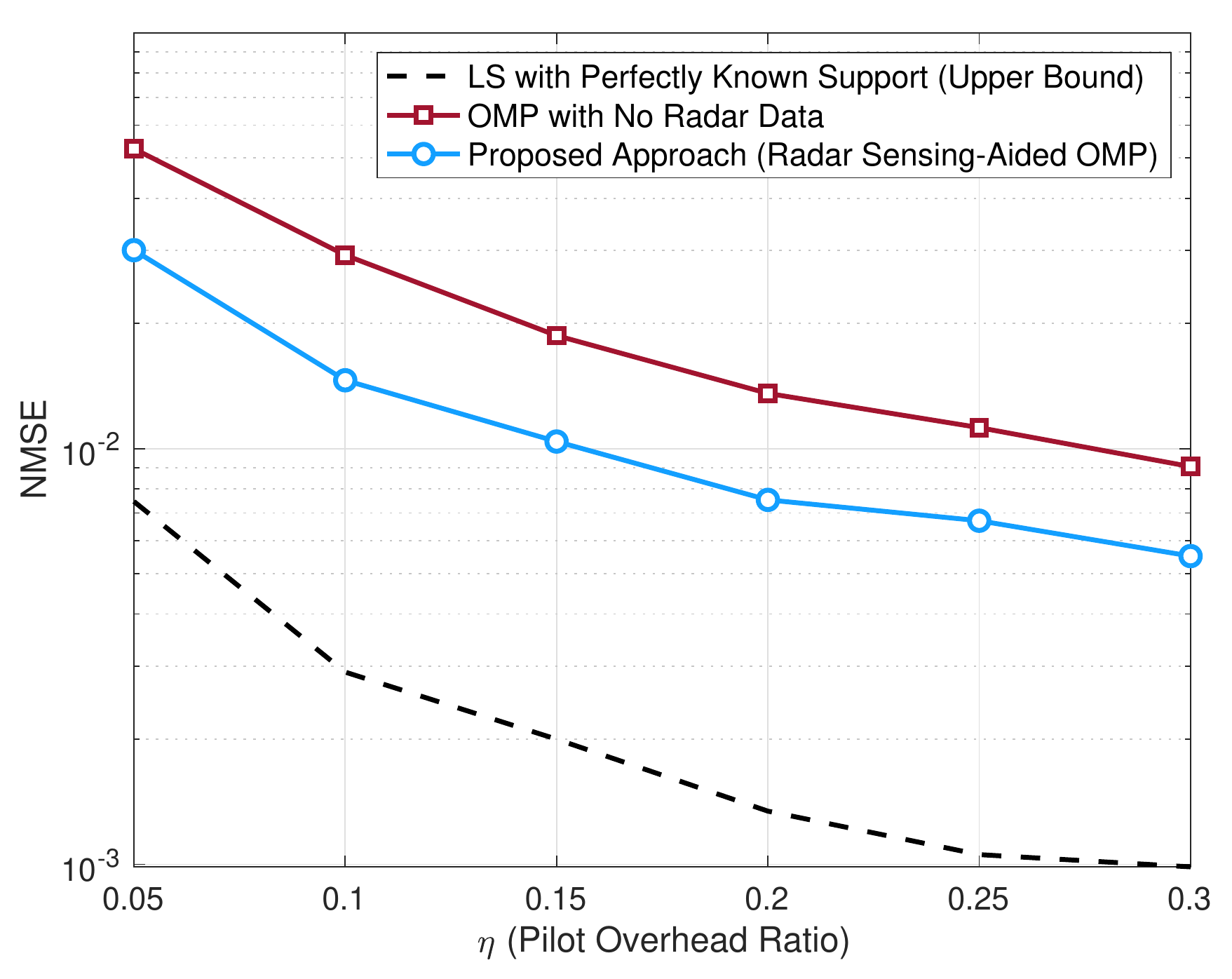}
		\caption{This figure shows the NMSE performance with different $\eta$ (pilot overhead ratio). The BS has $32$ antennas. The SNR is set to $10$dB.}
		\label{fig:eta}
	\end{figure}

	\subsection{Does Sensing Reduce the Pilot Overhead?}
	In Fig. \ref{fig:eta}, we present the NMSE performance comparison under various pilot overhead ratios $\eta$. The SNR here is set to $10$dB. It can be observed that the NMSE decreases when more pilot signals are used for channel estimation, which is expected. The proposed approach consistently outperforms the  conventionaly OMP (angle domain) for all pilot overhead values. Particularly, the proposed approach achieves similar NMSE performance using $\eta=0.15$ compared to the ``OMP (angle domain)" using $\eta=0.3$. \textbf{This indicates a $\mathbf{50\%}$ reduction in the pilot overhead.} This result captures the prospect of utilizing radar sensing to reduce channel estimation overhead. Note that the LS with perfectly known support (which can act as an upper bound) has much better performance than the other approaches. This indicates that there is still a clear room for enhancements on the proposed approach that attempts to estimate the support using radar sensing, which motivates future research.

	\section{Conclusions}\label{Conclusion}
	In this paper, we investigated the downlink channel estimation problem for massive MIMO-OTFS systems and developed a novel approach that leverages the radar sensing information at the basestation to aid the OTFS channel estimation task. This is particularly motivated by the integratration of sensing and communication in future wireless systems and by the direct relationship between the delay-Doppler channel and the sensing infomration  (such as location/velocity/direction) about the mobile user/scatterers in the environment. We formulated the delay-Doppler channel estimation problem as a sparse recovery problem and utilized radar sensing  to aid the compressive sensing solution. Using accurate 3D ray tracing, we constructed an evaluation platform with co-existing communication and radar sensing data and used it to assess the performance of the proposed solution. The results showed that the proposed sensing-aided OTFS channel estimation approach consistently outperforms the conventional OMP in terms of both the channel estimation NMSE and the required pilot overhead, highlighting a promising approach for future OTFS massive MIMO systems.



\begin{thebibliography}{10}
	\providecommand{\url}[1]{#1}
	\csname url@samestyle\endcsname
	\providecommand{\newblock}{\relax}
	\providecommand{\bibinfo}[2]{#2}
	\providecommand{\BIBentrySTDinterwordspacing}{\spaceskip=0pt\relax}
	\providecommand{\BIBentryALTinterwordstretchfactor}{4}
	\providecommand{\BIBentryALTinterwordspacing}{\spaceskip=\fontdimen2\font plus
		\BIBentryALTinterwordstretchfactor\fontdimen3\font minus
		\fontdimen4\font\relax}
	\providecommand{\BIBforeignlanguage}[2]{{%
			\expandafter\ifx\csname l@#1\endcsname\relax
			\typeout{** WARNING: IEEEtran.bst: No hyphenation pattern has been}%
			\typeout{** loaded for the language `#1'. Using the pattern for}%
			\typeout{** the default language instead.}%
			\else
			\language=\csname l@#1\endcsname
			\fi
			#2}}
	\providecommand{\BIBdecl}{\relax}
	\BIBdecl
	
	\bibitem{Hadani2017}
	R.~Hadani, S.~Rakib, M.~Tsatsanis, A.~Monk, A.~J. Goldsmith, A.~F. Molisch, and
	R.~Calderbank, ``Orthogonal time frequency space modulation,'' in \emph{2017
		IEEE Wireless Communications and Networking Conference (WCNC)}, 2017, pp.
	1--6.
	
	\bibitem{Sayeed2021}
	A.~M. Sayeed, ``How is time frequency space modulation related to short time
	fourier signaling?'' in \emph{2021 IEEE Global Communications Conference
		(GLOBECOM)}, 2021, pp. 1--6.
	
	\bibitem{shen2019channel}
	W.~Shen, L.~Dai, J.~An, P.~Fan, and R.~W. Heath, ``Channel estimation for
	orthogonal time frequency space ({OTFS}) massive mimo,'' \emph{IEEE
		Transactions on Signal Processing}, vol.~67, no.~16, pp. 4204--4217, 2019.
	
	\bibitem{guo2020convolutional}
	J.~Guo, C.-K. Wen, S.~Jin, and G.~Y. Li, ``Convolutional neural network-based
	multiple-rate compressive sensing for massive mimo {CSI} feedback: Design,
	simulation, and analysis,'' \emph{IEEE Transactions on Wireless
		Communications}, vol.~19, no.~4, pp. 2827--2840, 2020.
	
	\bibitem{rasheed2020sparse}
	O.~K. Rasheed, G.~Surabhi, and A.~Chockalingam, ``Sparse delay-{Doppler}
	channel estimation in rapidly time-varying channels for multiuser {OTFS} on
	the uplink,'' in \emph{2020 IEEE 91st Vehicular Technology Conference
		(VTC2020-Spring)}.\hskip 1em plus 0.5em minus 0.4em\relax IEEE, 2020, pp.
	1--5.
	
	\bibitem{zhang20182d}
	M.~Zhang, F.~Wang, X.~Yuan, and L.~Chen, ``{2D} structured turbo compressed
	sensing for channel estimation in {OTFS} systems,'' in \emph{2018 IEEE
		International Conference on Communication Systems (ICCS)}.\hskip 1em plus
	0.5em minus 0.4em\relax IEEE, 2018, pp. 45--49.
	
	\bibitem{shi2021deterministic}
	D.~Shi, W.~Wang, L.~You, X.~Song, Y.~Hong, X.~Gao, and G.~Fettweis,
	``Deterministic pilot design and channel estimation for downlink massive
	{MIMO-OTFS} systems in presence of the fractional {Doppler},'' \emph{IEEE
		Transactions on Wireless Communications}, vol.~20, no.~11, pp. 7151--7165,
	2021.
	
	\bibitem{Demirhan_mgazine_radar}
	\BIBentryALTinterwordspacing
	U.~Demirhan and A.~Alkhateeb, ``Integrated sensing and communication for {6G}:
	Ten key machine learning roles,'' 2022. [Online]. Available:
	\url{https://arxiv.org/abs/2208.02157}
	\BIBentrySTDinterwordspacing
	
	\bibitem{liu2020joint}
	F.~Liu, C.~Masouros, A.~P. Petropulu, H.~Griffiths, and L.~Hanzo, ``Joint radar
	and communication design: Applications, state-of-the-art, and the road
	ahead,'' \emph{IEEE Transactions on Communications}, vol.~68, no.~6, pp.
	3834--3862, 2020.
	
	\bibitem{Demirhan2022}
	U.~Demirhan and A.~Alkhateeb, ``Radar aided 6g beam prediction: Deep learning
	algorithms and real-world demonstration,'' in \emph{2022 IEEE Wireless
		Communications and Networking Conference (WCNC)}, 2022, pp. 2655--2660.
	
	\bibitem{Zhang2021radar}
	J.~A. Zhang, F.~Liu, C.~Masouros, R.~W. Heath, Z.~Feng, L.~Zheng, and
	A.~Petropulu, ``An overview of signal processing techniques for joint
	communication and radar sensing,'' \emph{IEEE Journal of Selected Topics in
		Signal Processing}, vol.~15, no.~6, pp. 1295--1315, 2021.
	
	\bibitem{Kumari18}
	P.~Kumari, J.~Choi, N.~González-Prelcic, and R.~W. Heath, ``Ieee
	802.11ad-based radar: An approach to joint vehicular communication-radar
	system,'' \emph{IEEE Transactions on Vehicular Technology}, vol.~67, no.~4,
	pp. 3012--3027, 2018.
	
	\bibitem{wei2021transmitter}
	Z.~Wei, W.~Yuan, S.~Li, J.~Yuan, and D.~W.~K. Ng, ``Transmitter and receiver
	window designs for orthogonal time-frequency space modulation,'' \emph{IEEE
		transactions on communications}, vol.~69, no.~4, pp. 2207--2223, 2021.
	
	\bibitem{raviteja2018interference}
	P.~Raviteja, K.~T. Phan, Y.~Hong, and E.~Viterbo, ``{Interference cancellation
		and iterative detection for orthogonal time frequency space modulation},''
	\emph{IEEE transactions on wireless communications}, vol.~17, no.~10, pp.
	6501--6515, 2018.
	
	\bibitem{asuzu2018more}
	P.~Asuzu and C.~Thompson, ``A more exact linear fmcw radar signal model for
	simultaneous range-velocity estimation,'' in \emph{2018 IEEE Radar Conference
		(RadarConf18)}.\hskip 1em plus 0.5em minus 0.4em\relax IEEE, 2018, pp.
	0007--0011.
	
	\bibitem{3GPPchannel}
	{3GPP TR 25.996 version 11.0.0 Release 11}, ``{Universal Mobile
		Telecommunications System (UMTS); Spatial channel model for Multiple Input
		Multiple Output (MIMO) simulations},'' {}, Tech. Rep., 2012.
	
	\bibitem{BP}
	S.~S. Chen, D.~L. Donoho, and M.~A. Saunders, ``Atomic decomposition by basis
	pursuit,'' \emph{SIAM review}, vol.~43, no.~1, pp. 129--159, 2001.
	
	\bibitem{MP}
	S.~G. Mallat and Z.~Zhang, ``Matching pursuits with time-frequency
	dictionaries,'' \emph{IEEE Transactions on signal processing}, vol.~41,
	no.~12, pp. 3397--3415, 1993.
	
	\bibitem{rappaport2019wireless}
	T.~S. Rappaport, Y.~Xing, O.~Kanhere, S.~Ju, A.~Madanayake, S.~Mandal,
	A.~Alkhateeb, and G.~C. Trichopoulos, ``Wireless communications and
	applications above 100 {GHz}: Opportunities and challenges for {6G} and
	beyond,'' \emph{IEEE access}, vol.~7, pp. 78\,729--78\,757, 2019.
	
	\bibitem{feng2020joint}
	Z.~Feng, Z.~Fang, Z.~Wei, X.~Chen, Z.~Quan, and D.~Ji, ``Joint radar and
	communication: A survey,'' \emph{China Communications}, vol.~17, no.~1, pp.
	1--27, 2020.
	
	\bibitem{rohling1983radar}
	H.~Rohling, ``Radar {CFAR} thresholding in clutter and multiple target
	situations,'' \emph{IEEE transactions on aerospace and electronic systems},
	no.~4, pp. 608--621, 1983.
	
	\bibitem{tropp2007signal}
	J.~A. Tropp and A.~C. Gilbert, ``{Signal recovery from random measurements via
		orthogonal matching pursuit},'' \emph{IEEE Transactions on information
		theory}, vol.~53, no.~12, pp. 4655--4666, 2007.
	
\end{thebibliography}
\end{document}